\documentclass{elsart}
\usepackage{epsfig,graphicx, amssymb}
\begin{document}
\runauthor{Litak et al.}

\begin{frontmatter}
\title{Estimation of a Noise Level Using Coarse-Grained Entropy of
Experimental
Time Series of Internal Pressure in a Combustion Engine}
\author[Lublin1]{Grzegorz Litak\thanksref{E-mail},}
\author[Trieste]{Rodolfo Taccani,}
\author[Trieste]{Robert Radu,}
\author[Warszawa]{Krzysztof Urbanowicz,}
\author[Warszawa]{Janusz A. Ho\l{}yst,}
\author[Lublin2]{Miros\l{}aw Wendeker}
\author[Trieste]{and Alessandro Giadrossi}

\address[Lublin1]{Department of Mechanics, Technical University of
Lublin,
Nadbystrzycka 36, PL-20-618 Lublin, Poland}


\address[Warszawa]{Faculty of Physics, Warsaw University of Technology,
Koszykowa 75, PL-00-662, Warsaw, Poland}

\address[Trieste]{Department of Energetics, University of Trieste
Via A. Valerio 10, I--34127 Trieste, Italy}

\address[Lublin2]{Department of Combustion Engines, Technical University
of
Lublin,
Nadbystrzycka 36, PL-20-618 Lublin, Poland}

\thanks[E-mail]{Fax: +48-815250808; E-mail:
g.litak@pollub.pl}

\begin{abstract}
We report our results on non-periodic
experimental time series of pressure in a single cylinder spark
ignition
engine.
The experiments were performed for different levels of loading.
We estimate the noise level  in internal pressure calculating the coarse-grained
entropy from variations of  maximal pressures in successive cycles.
The results show that  the dynamics of the
combustion is
a nonlinear multidimensional process mediated by noise.
Our results show that so defined level of noise in internal pressure 
is not monotonous function of loading. 
\end{abstract}
\end{frontmatter}

\section{Introduction}
The cycle-to-cycle combustion variability has been a subject of
interest for many years
\cite{Hey88,Dai88,Foa93,Che94,Hu96,Rob97,Daw96,Daw98,Gre99,Wen03,Wen04,Kam04}. 
It is 
possible
that many different disturbances influence the process making it
stochastic but in case of high sensitivity  on the process
conditions one should also consider appearance of chaotic behaviour
\cite{Dai88,Foa93,Che94,Daw96,Daw98,Gre99,Wen03,Wen04}. The pressure 
signal can be used
to control the engine combustion \cite{Leo99}. Its  
variations 
might
originate from a complex dynamics leading, presumably with some
stochastic component, to non-periodic behaviour. Thus, the crucial
problem  is to understand the nonlinear dynamics of the process
observing the internal pressure inside the cylinder. In the
present note we discuss experimental results of a direct
measurement of internal pressure and analyse its time series using
the coarse grained entropy method \cite{Urb03}.
The method  has been already used successively
by Kaminski et al. \cite{Kam04}
to examine the dynamics of combustion in a four cylinder combustion engine to examine the
influence of an spark advance angle on cycle-to-cycle variations of heat 
release. 
In this paper we present 
analysis of pressure concerned to a single cylinder engine.   
Experimental works
have been performed in the Engine Laboratory at University of Trieste. 

The paper is organised as follows. After the present introduction we will discuss
experimental facilities in Sec. 2. Following Sec. 3 is devoted to
 analysis of pressure time series.  In this section we show the experimental 
results and perform their initial analysis.
In Sec. 4 we present the methodology of noise level estimation and we finally apply 
the coarse-grained entropy method to
internal pressure time series. 
At the end we provide
conclusions in the last Sec. 5.

\begin{figure}

\begin{center}
\epsfig{file=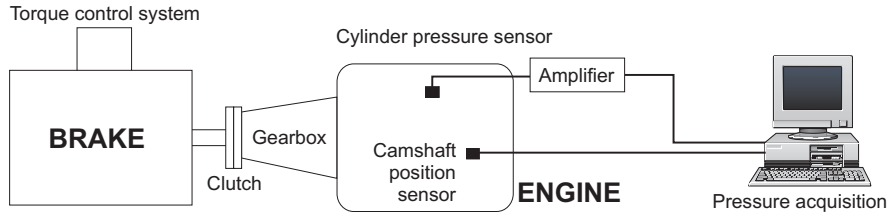,width=12.0cm,angle=0}
\end{center}

\caption{\label{rys1}
Schematics of the engine pressure measurement and control setup.
}
\vspace{0.5cm}
\end{figure}

\section{Experimental Facilities}
Tests were carried out on a SI - four stroke, single cylinder,
Aprilia/Rotax engine. The instrumentation which was used during
our test is schematically presented
 in Fig. \ref{rys1}.

The pressure sensor Kistler 6053C, was preferred for its small dimensions.
Mounting
 the pressure sensor was a complex task because of the cylinder head profile.
The sensor was placed in the relatively small available space, which remains
between
 the five valves areas. It was centrally positioned, having the active part very
 closed to the spark plug electrodes. 
As the upper end of the sensor, stays under 
the cylinder
head
 cover, a protective tube for the connecting cable was mounted.

  The encoder (Lika C58L), due to the engine configuration, was fixed on 
the camshaft.

This
 solution  generates complications regarding signal processing. Because of 
the transmission
 chain, there  is no rigid connection between the engine crankshaft and 
the
encoder shaft,
therefore
  synchronisation problems may appear. To overcome this, before each 
testing session, motored
  engine tests were accomplished, to determine the shifting between the 
encoder trigger and
 pressure maximum peak.

The reference level for the cylinder pressure was established using the intake pressure
 signal, acquired with an inductive pressure sensor. Before each test session, this
sensor
was
 calibrated using a simple system containing: a piston pump which creates the
 calibration pressure, a high precision manometer, a signal amplifier which amplifies the
 sensor output and a voltmeter \cite{Gia99}.

\begin{figure}
\hspace{3.0cm}
 \epsfig{file=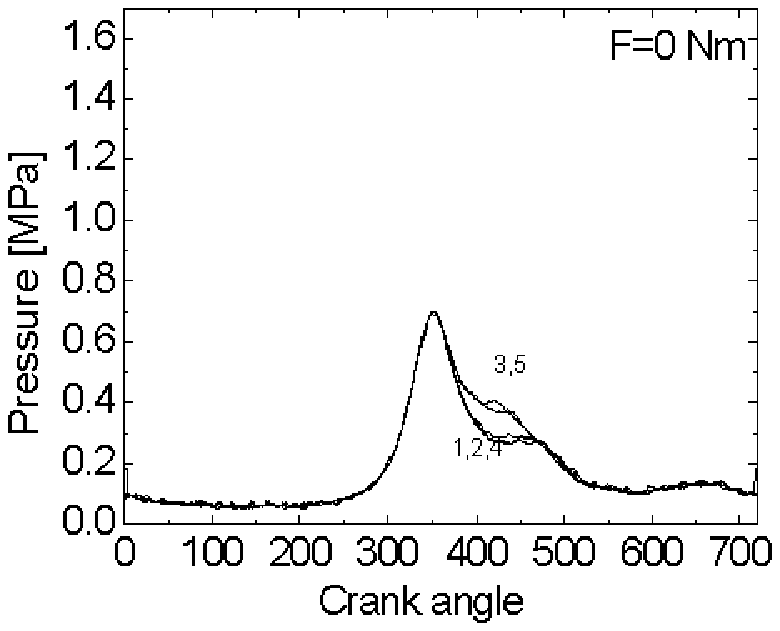,width=8.0cm,angle=-0}

\hspace{3.0cm}
\epsfig{file=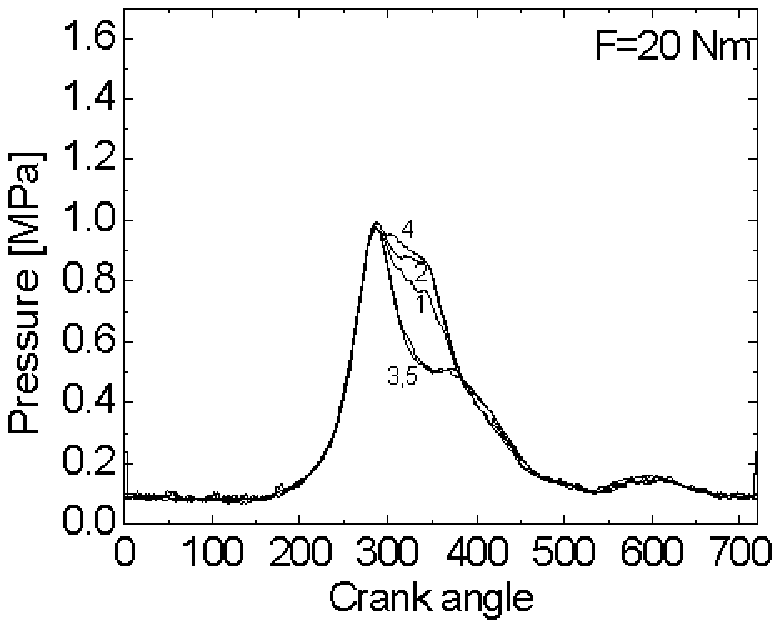,width=8.0cm,angle=-0}

\hspace{3.0cm}
\epsfig{file=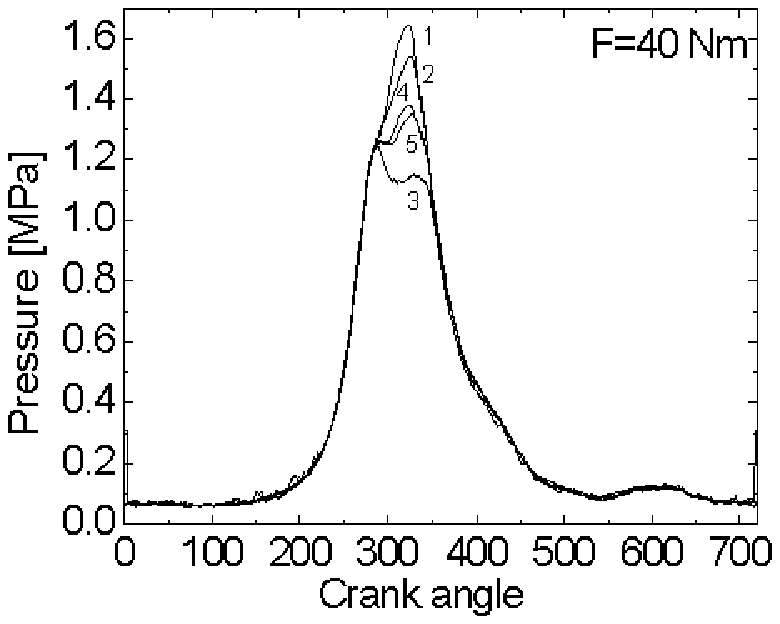,width=8.0cm,angle=-0}

~ \caption{\label{rys2} 
Pressure in first five cycles of measured engine work against  the crank angle $\phi$
for   different torque loadings $F=0$, 20 and 40 Nm at 4000 rpm. Numbers in figures 1-5 
correspond 
to pressure curves in successive cycles, respectively.
 } \vspace{1cm}
\end{figure}

\begin{table}

\caption{\label{tableone} Definitions of variables and symbols used in the
paper.
}
\vspace{1cm}
\hspace{2cm}
\begin{tabular}{c|c}
 \hline
number of measured pressure consecutive in averaging procedure & r\\
internal cylinder pressure & $P_i$ \\
maximum pressure in cycle & $P^{max}_i$ or $P_{max}$ \\
average of maximum pressure series & $\overline{P_{max}}$ \\
 maximum pressure vector in the embedding space  & {\bf P}$^{max}_i$ \\
Heaviside step function
 &  $\Theta(z)$ \\
embedding time delay in cycles&  $m$ \\
cycle number & $i,j$ \\
embedding dimension & n \\
number of considered points in time series & $N$ \\
loading torque & $F$ \\
crank angle & $\phi \in [0,720^o]$ \\
threshold & $\varepsilon$ \\
correlation integral & $C^n$ \\
coarse-grained correlation integral & $C^n (\varepsilon)$ \\
correlation entropy & $K_2$ \\
coarse-grained correlation entropy & $K_2(\varepsilon)$ \\
calculated from time series coarse-grained entropy & $K_{noisy}$ \\
correlation dimension & $D_2$ \\
Noise-to-Signal ratio & NTS \\
standard deviation of data & ${\sigma}_{DATA}$\\
error function & Erf$(z)$ \\
fitting parameters & $\kappa$ , $a$, $b$ \\
standard deviation of noise & $\sigma$ \\
top dead center & TDC \\
\hline
\end{tabular}
\vspace{5cm}
~
\end{table}

\section{Time Series Analysis}

Using our experimental standing we have measured internal pressure
time series for 1000 cycles measuring 6000 points per cycle. 
In Fig. 2 we have plotted pressure against the crank angle  $\phi$
for first five consecutive cycles of measured engine work against  
the crank angle $\phi$ for  different torque loadings $F=0$, 20 and 40 Nm. 
Note that with larger loads the peak pressure is higher and 
the combustion zone concentrates near the TDC.
Variations of pressure reflects the efficiency of
combustion process
in successive cycles. 
Such a phenomenon is associated with the burned fuel fluctuations and
influence negatively on its
consumption. The cyclic fluctuations of pressure in combustion process
can be
described by its maximal value
$P^{max}$ in each cycle. 
Interestingly, as it was proposed in Ref. \cite{Leo99}, it is
possible to use the maximal values of pressure $P^{max}_i$
involved in control process to achieve stable combustion. The
method of peak-to-peak analysis appeared to be useful to analyse
some kind of dynamical systems \cite{Pic03}, as it is in our case.
This peak-to-peak variation is our starting point for
further analysis by means of a
Coarse-Grained Entropy and an autocorrelation function.

\begin{figure}

\vspace{-1.5cm}
\epsfig{file=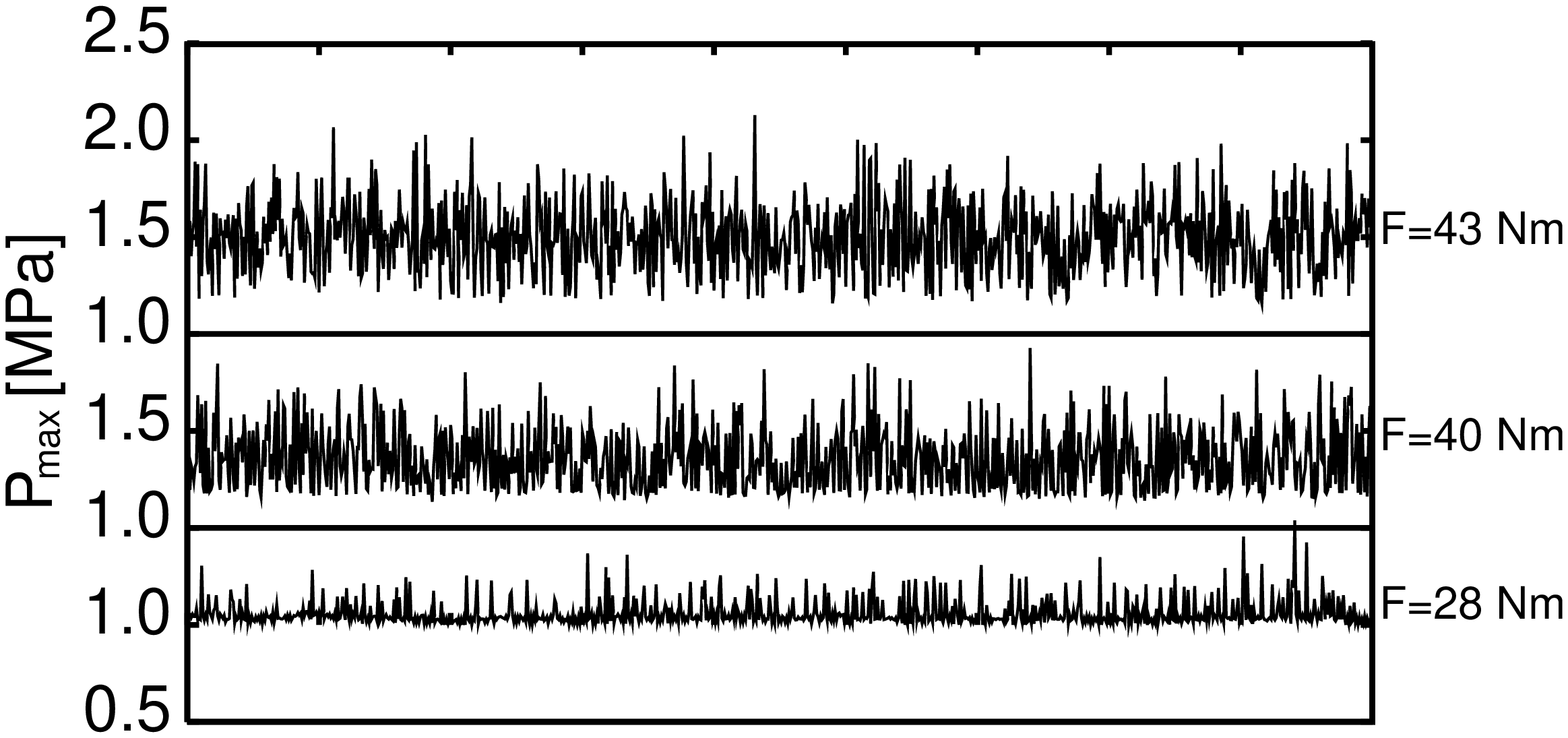,width=6.0cm,angle=-90}

\epsfig{file=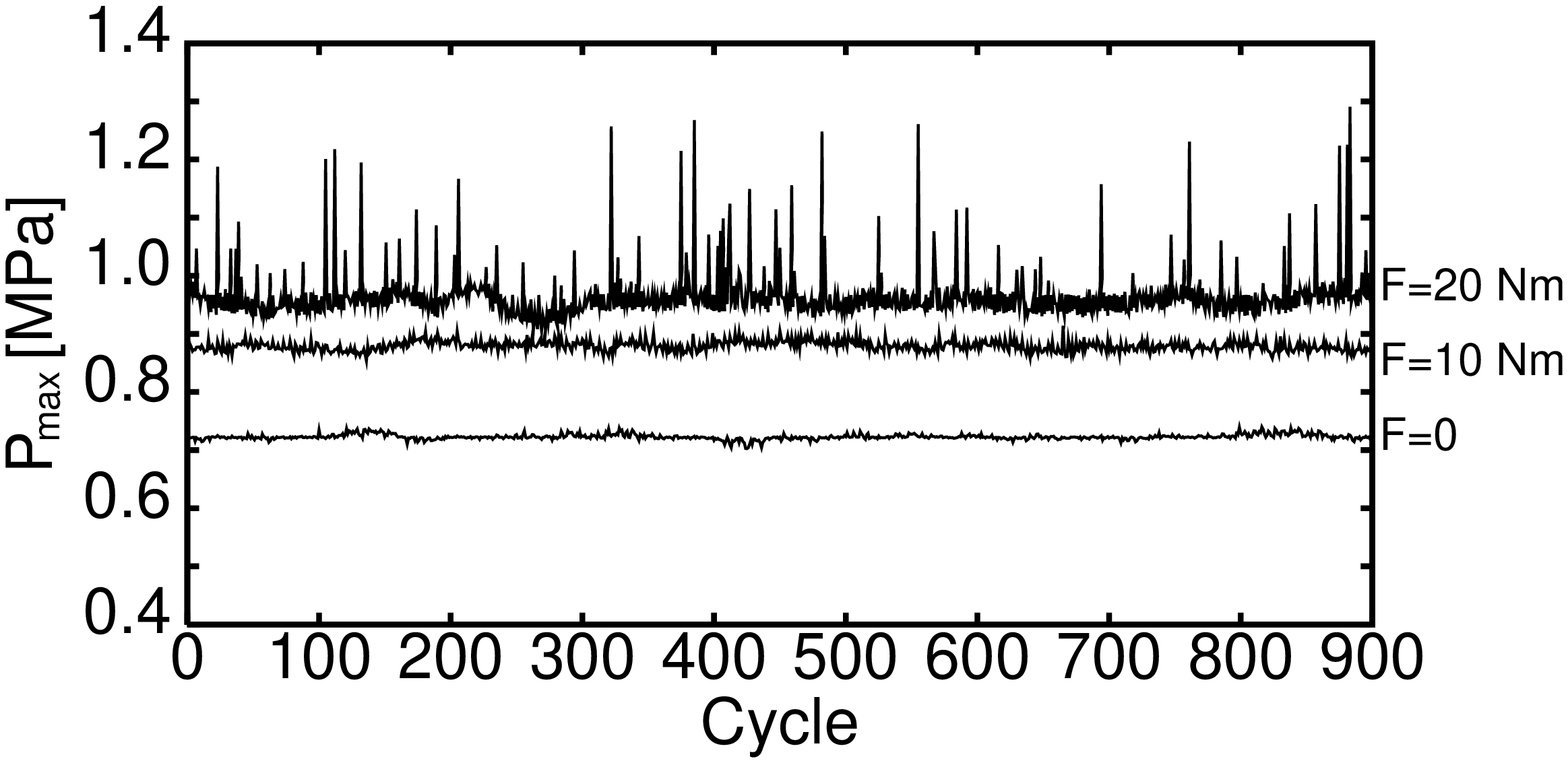,width=6.0cm,angle=-90}

\caption{\label{rys4} Cycle-to-cycle variations of maximal
pressure  values $P_{max}$ of 900 cycles at different loads $F=0$, 10, 20, 28, 
40,
43 Nm,
respectively. Note different scales in lower and upper panels. }
\end{figure}

\begin{figure}
\hspace{1.5cm}
\epsfig{file=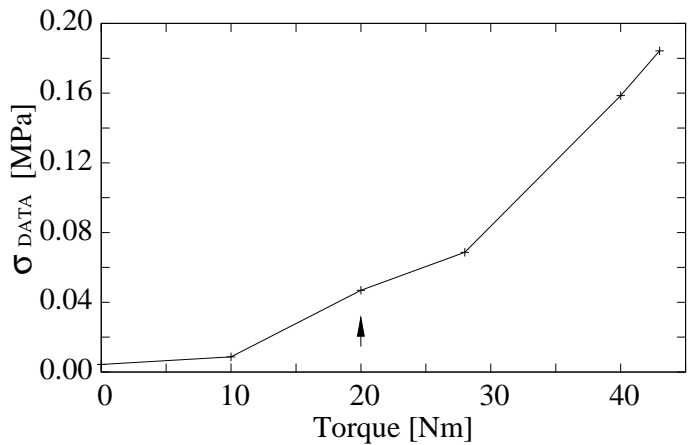,width=11cm,angle=-0}
\caption{
Standard deviation ${\large \sigma}_{DATA}$ of $P_{max}$ signals versus an applied 
torque $F$. 
Note, the
arrow at
$F=20$ Nm indicates the first sudden grow of ${\large \sigma}_{DATA}$ for relatively low 
loading.
}
\end{figure}

To show its cycle-to-cycle changes we
have plotted $P^{max}$ against
successive
cycles $i$ in Fig. \ref{rys4}. One should note that both, average value of 
pressure  $\overline{P_{max}}$
as well as  
its standard deviation  
 ${\large \sigma}_{DATA}$ (Fig. 4) increase with growing torque $F$. 
Interestingly, for about $F=20$ we observe the sudden growth of  ${\large 
\sigma}_{DATA}$. 
This is a result of some
diversity  of a maximal pressure origin  as pressure itself is a complex product of 
compression 
and combustion phenomena. Clearly, one can see the intermittent behaviour of $P_{max}$  
(Fig. 3  for $F=20$ and 28 Nm)
can be explained in these terms. Lean combustion where the maximum of pressure is 
produced largely by piston compression with possible ignition missing corresponds to 
lower level of $P_{max}$ (Fig. 3)
while effective combustion happens from time to time corresponds to spikes in Fig. 3.
Similar effects can be seen for $F=28$ and $F=40$ but not for $F=43$ where
the distribution of $P_{max}$ is going back to a Gaussian type.

\section{Noise Level Estimation}

In the $n$ dimensional
   embedding space \cite{Kan97} the state is determined by a
multidimensional vector
\begin{equation}
\label{eq1}
{\bf P}^{max}=
\{P^{max}_i,P^{max}_{i+m},P^{max}_{i+2m},...,P^{max}_{i+(n-1)m} \},
\end{equation}
where $m$ denotes the embedding delay in terms of cycles. The
correlation integral calculated in the embedding space can be
defined as
 \cite{Paw87,Gra83}

\begin{equation}
\label{eq2}
C^n(\varepsilon) =
 \frac{1}{N^2} \sum_{i}^{N} \sum_{j \neq i}^{N} \Theta ( \varepsilon
-|| \mathbf{P}^{max}_{i}-\mathbf{P}^{max}_{j} ||),
\end{equation}
where $N$ is the number of considered points corresponding to
pressure peaks in cycles and $\Theta$ is the Heaviside step  
function. For simplicity we use maximum norm. The correlation
integral $C^n(\varepsilon)$ is related to the correlation entropy
$K_2(\varepsilon)$ and the system correlation dimension $D_2$ by 
the following formula \cite{Paw87,Gra83}

\begin{figure}

\hspace{1cm}
\epsfig{file= 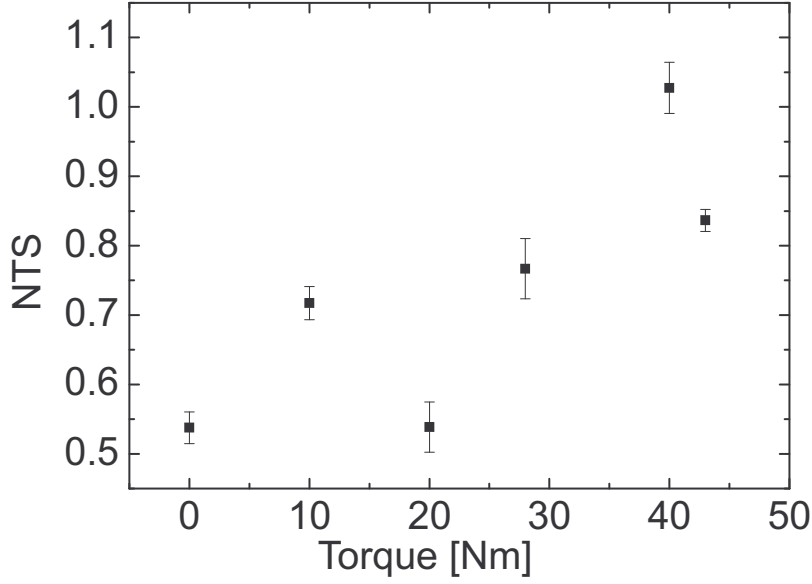,width=8.0cm,angle=-90}

\caption{\label{rys5}
Level of noise estimated for different loads (Torque).
}
\vspace{0.5cm}
\end{figure}

\begin{equation}
\label{eq4}
 \lim_{n \rightarrow \infty} C^n(\varepsilon)=D_2
\ln \varepsilon -nmK_2(\varepsilon).
\end{equation}

The coarse-grained correlation entropy can be now be calculated as

\begin{equation}
\label{eq3}
K_2(\varepsilon)=  \lim_{n \rightarrow \infty} \ln \frac{C^n
(\varepsilon)}{C^{n+1}
(\varepsilon)} \approx - \frac{ {\rm d} \ln C^n (\varepsilon)}{{\rm d}
n}.
\end{equation}

In such a case the correlation entropy is defined in the limit of
a small threshold $\varepsilon$.

In presence of noise described by the standard deviation $\sigma$
of ${\bf P}^{max}_i$ time series,
the observed coarse-grained entropy $K_{noisy}$
\cite{Urb03} can be written as
\begin{eqnarray}
\label{eq5}
K_{noisy}(\varepsilon) &=& -\frac{1}{m} {\rm g}\left(\frac{\varepsilon}{2
\sigma} \right)
 \ln \varepsilon + \left[ \kappa +b \ln (1-a \varepsilon) \right]
 \nonumber
\\
&\times& \left( 1 +\sqrt{\pi} \frac{\sqrt{ \varepsilon^2/3 +2
\sigma^2} - \varepsilon/\sqrt{3}}{\varepsilon}   \right).
\end{eqnarray}

\noindent Function g$(z)$, present in the above formula, reads   
\begin{equation}
\label{eq6}
{\rm g}(z)=\frac{2}{\pi} \frac{z {\rm e}^{-z^2}}{{\rm Erf}(z)},
\end{equation}

where Erf$(.)$ is the Error Function. The parameters $\kappa$,
$a$, $b$ as well as $\sigma$ are unconstrained. They should be
fitted in Eq. (\ref{eq5}) to mimic the observed noisy entropy
calculated from the experimental data.

\begin{figure}

\hspace{1.5cm}
\epsfig{file= 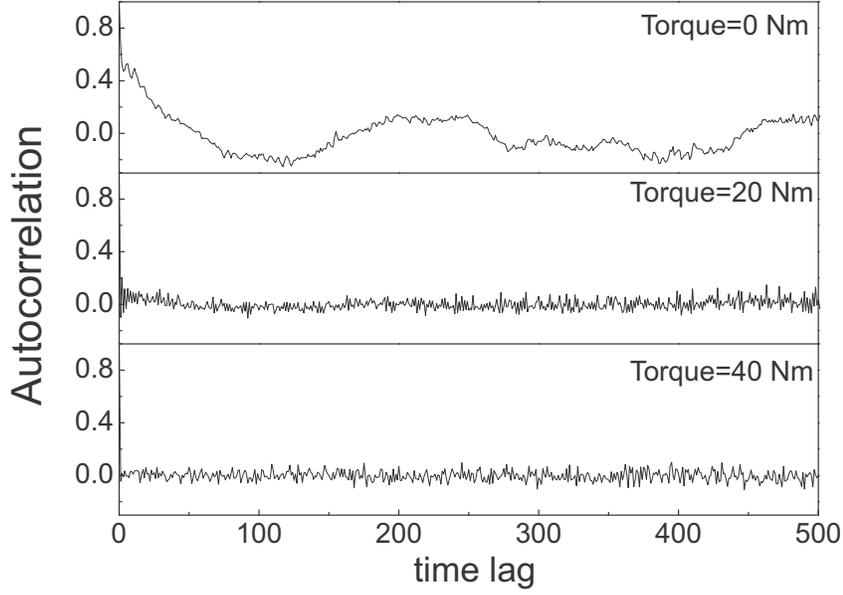,width=8cm,angle=-90}

\caption{\label{rys6}
Autocorrelation function of the maximal pressure signal.
}
\vspace{1cm}   
\end{figure}

Using the above procedure in analysing the time series of pressure
peaks we have calculated the coarse-grained correlation entropy
for different loading levels ($F=10$, 20, 28, 40, 43Nm).  Fitting it to
the last formula (Eq. \ref{eq5}) it was possible to estimate the
noise level $\sigma$. In Fig. \ref{rys5} we present the obtained
results with the help of NTS parameter defined as follows
\begin{equation}
{\rm NTS}=\frac{\sigma}{\sigma_{DATA}},
\end{equation}
where $\sigma_{DATA}$ is the standard deviation of maximum pressure signal data (Fig. 
4). 
The error on Fig.5 comes from the fitting procedure of Eq.(5) to
observed coarse-grained correlation entropy.
 Note,
the leading tendency is that for higher loading the noise is
larger, what is expected for most of the engines with lean combustion. However the
cases of $F=20$ and 43Nm could be the exception of this rule. 
At $F=20$ we simultaneously observe the first sudden increase of ${\large \sigma}_{DATA}$.
It is natural that the overall effect (Eq. 7) will cause decreasing of  NTS.
Interestingly, the nodal or very low
loading $F=0$  signal (of $P_{max}$) appears to be influenced more by 
a compression phenomenon mediated by combustion process (Fig. 2a) than  
combustion itself. In this case $P_{max}$ is 
highly correlated. 
In Fig.
\ref{rys6} we show the autocorrelation function for different torque loading $F=0$, 20 
and
40Nm. Note that for the lowest loading ($F=0$) the autocorrelation function
does not decrease in contrast to the other cases.

On the other hand at  $F=43$ Nm (Fig. 5) we have to do with a different situation. 
In that case due 
to increasing 
loading
the fresh fuel feeding  is of higher level and 
cycles which misses of ignition do not happen any more.
On the other hand for  $F=40$ and 43 Nm we observe further growing  of the signal 
square deviation $\sigma_{DATA}$
with an increasing  load.
In such conditions 
 it is 
not a surprise that 
in such a case we also observe 
a lower level of  NTS.

\section{Summary and Conclusions}

We applied the Coarse-Grained Entropy Method for the
experimental data estimating the noise-level. 
This method enable to distinguish determined signal, including chaos with its short time 
prediction scale, and random noise.
Analysing our
experimental data we decided to use maximal values of pressure
$P^{max}$. 
In all considered cases we have found a relatively high noise level.
However it has  appeared that larger loading  led to
more noisy system.  For larger power the engine 
needs more fuel. Increase of
fluctuation in combustion can  happen if the engine work is 
optimised for the lowest fuel consumption or the exhaust gas 
toxicity. Another interesting point, discussed in the paper, was connected with a 
missing ignition effect. 
In fact such effects are present for small loading (Figs. 2a-b) but influence strongly 
$P_{max}$ distribution only if loading is large enough.

We have also found that the pressure signal $P^{max}$ was more
correlated (Fig. 6) for small loading. 
Note, the noise level obtained for a correlated 
signal case 
by our method can be underestimated by a few percents.    
A noise  component for small loading  
can be also related to  a well known problem of idle speed engine
control difficulties \cite{Wen03a,Bad01}.

\section*{Acknowledgements}
Three of authors (GL, KU, JAH) would like to thank
Max Planck Institute for
Physics
of Complex
Systems in Dresden for hospitality.
During their stay in Dresden an important
part of data analysis has been performed.
KU is grateful to DAAD for a financial support.

\end{document}